\documentclass[12pt,a4page]{amsart} 
\usepackage{listings}
\usepackage{color}
\usepackage{amsaddr}
\usepackage{hyperref}

\definecolor{dkgreen}{rgb}{0,0.6,0}
\definecolor{gray}{rgb}{0.5,0.5,0.5}
\definecolor{mauve}{rgb}{0.58,0,0.82}
\calclayout

\def\EE{{\mathbb E}}

\def\ZZZ{{\mathbf Z}}

\def\AAA{{\mathbf A}}

\def\PPP{{\mathbf P}}

\def\RR{{\mathbb R}}

\title{\MakeLowercase{\emph{rlsm}}: R package for least squares Monte Carlo}
\author{Jeremy Yee}
\email{jeremyyee@outlook.com.au}

\begin{document}

\maketitle

\begin{abstract}
  This short paper briefly describes the implementation of the least
  squares Monte Carlo method in the \emph{rlsm} package. This package
  provides users with an easy manner to experiment with the large
  amount of \emph{R} regression tools on any regression basis and
  reward functions. This package also computes lower and upper bounds
  for the true value function via duality methods.
\end{abstract}


\section{Introduction}

The popularity of the least squares Monte Carlo method
\cite{carriere1996,tsitsiklis_vanroy1999,longstaff_schwartz2001} has
been largely spurred on by its applications in finance and real
options valuation. This method uses statistical regression to
represent the continuation value functions in the Bellman recursion as
a linear combination of selected feature functions. Unlike tradition
approaches such as finite differences and tree methods, least squares
Monte Carlo is largely independent of the dimension of the state space
and so mostly avoids the so-called curse of dimensionality that is
common in dynamic programming problems. A rigiourous treatment of this
method is beyond the scope of this paper and so the reader is referred
to the work done by \cite{belomestny_etal2010} and the references
contained within for a more comprehensive analysis of this approach
and its convergence properties.  In its typical form, least squares
Monte Carlo employs linear least squares regression. However, as shown
by \cite{tompaidis_yang2014}, other regression approaches may be more
appropriate such as Ridge regression or Least Absolute Shrinkage and
Selection Operator (LASSO).  Numerous regression approaches have been
honed by statisticians and coded into the \emph{R} statistical
language \cite{R}. The aim of the \emph{rlsm} package is to allow
least squares Monte Carlo users to access the large amount of
regression tools developed by the \emph{R} community. To the author's
knowledge, this has not been done before in an \emph{R} package.  Note
that this package only focuses on global regression methods (as
opposed to local methods such as nearest neighbours). The \emph{rlsm}
\cite{rlsm} package is able to handle any specification of the
regression basis and reward functions. In addition, this package also
implements the dual approach studied by \cite{
  andersen_broadie2004,haugh_kogan2004,rogers2007,brown_etal2010} to
construct lower and upper bound for the unknown value function using a
pathwise dynamic programming approach. The computational effort is
done at \emph{C++} level via \emph{Rcpp} \cite{rcpp}.  Before
proceeding, let us make a key point. This paper neglects some
mathematical rigour in exchange for brevity. However, references are
provided for the interested reader. The paper is structured as
follows. The next section gives the problem setting. Section 3 gives a
description of the least squares Monte Carlo algorithm. Section 4
details the construction of the lower and upper bounds. Section 5
demonstrates the usage of the \emph{rlsm} package on a simple optimal
stopping problem. Section 6 concludes this paper.

\section{Markov decision process}
\label{sec_problem}

Let $\mathbf{X} = \mathbf{P} \times \ZZZ$ represent our state space
and is the product of a finite set $\mathbf{P}$ and a subset of the
Euclidean space $\ZZZ\subseteq\mathbb{R}^d$.  At time
$t = 0, 1, \dots, T$, an action $a \in \mathbf{A}$ is chosen and
these actions influences the evolution of the stochastic process
$(X_t)_{t=0}^T := (P_t,Z_t)_{t=0}^T$. The discrete component
$(P_t)_{t=0}^T$ is a controlled Markov chain with transition
probabilities
$(\alpha_{p,p'}^a)_{p, p' \in \mathbf{P}, a \in \mathbf{A}}$, where
$\alpha_{p,p'}^a$ is the probability of moving from $p$ to $p'$ after
applying action $a$. The continuous component $(Z_t)_{t=0}^T$ evolves
according to $Z_{t+1} = f_{t+1}(W_{t+1}, Z_t)$ where
$(W_{t+1})_{t=0}^{T-1}$ are indepedent random variables and $f_{t+1}$
is a measurable function. At each time $t=0, \dots, T$ the decision
rule $\pi_{t}$ is given by a mapping
$\pi_{t}: \mathbf{X} \to \mathbf{A}$, prescribing at time $t$ an
action $\pi_{t}(p,z) \in \mathbf{A}$ for a given state
$(p, z) \in \mathbf{X}$. A sequence $\pi = (\pi_{t})_{t=0}^T$ of
decision rules is called a policy.  For each policy
$\pi = (\pi_{t})_{t=0}^{T}$, associate it with a so-called policy
value $v^{\pi}_{0}(p_0, z_{0})$ defined as the total expected
cumulative reward
\begin{equation} \label{value}
v^{\pi}_{0}(p_0, z_{0})=\mathbb{E}\left[\sum_{t=0}^{T-1}
r_{t}(P_t, Z^{}_{t}, \pi_{t}(X_t)) +  r_{T}(P_t, Z^{}_{t}) \right]
\end{equation}
where $r_t$ and $r_T$ are the reward and scrap functions,
respectively. A policy $\pi^{*}=(\pi^{*}_{t})_{t=0}^{T}$ is called
optimal if it maximizes the above expectation over all policies
$\pi \mapsto v^{\pi}_{0}(p, z)$. If an optimal policy exists, it
satisfies the Bellman recursion via
\begin{equation} \label{bellman}
\pi^{*}_{t}(p,z) = \arg\max_{a \in \mathbf{A}}\left\{r_{t}(p, z, a)+
\sum_{p'\in\mathbf{P}}
\alpha_{p,p'}^a\mathbb{E}[v^{*}_{t+1}(p',f_{t+1}(W_{t+1},z))] \right\}
\end{equation}
for $t = T-1,\dots, 0$. 

Note that the assumption that $(Z_t)_{t=0}^T$ is uncontrolled is a
simplyfing one since the least square Monte Carlo methods simulates a
number of scenarious from $(P_0,Z_0)$. Therefore, if the process
$(Z_t)_{t=0}^T$ is controlled, the use of a reference probability
measure and the corresponding densities is required to adjust the
conditional expectaions in \eqref{bellman} (see problem formulation in
Section 2 in \cite{belomestny_etal2010}).  This is difficult to
implement and not typically used in practice and so is not considered
in this paper and package.

\section{Least squares Monte Carlo (LSM)}

The goal of least squares Monte Carlo (LSM) is to express the conditional
expectations in \eqref{bellman} as a linear combination of basis
functions using values held by simulated paths. Suppose
$(Z_t(\omega_i))_{t=0}^T$ represents simulated trajectory $i$. At
terminal time $t=T$ and position $p$, the scrap
$\widetilde v_T(p,\omega_i) := r_T(p,Z_T(\omega_i))$ is realized for
each of the $n$ sample paths. Now at $t=T-1$, the values
$(\widetilde v_T(p,\omega_i))_{i=1}^n$ are then regressed on a chosen
regression basis constructed using $(Z_{T-1}(\omega_i))_{i=1}^n$ to
give an approximation of the conditional expectations which we will
denote by $\widetilde c_T^a(p',z)$ for $p'\in\PPP$.  Note that the
regression is performed using all the simulated paths.  Now for each
path $\omega_i$, determine fitted decision rule
$$
\widetilde\pi_{T-1}(p, \omega_i) := \arg\max_{a\in\AAA} \{r_{T-1}(p,
Z_{T-1}(\omega_i), a) + \sum_{p'\in\mathbf{P}} \alpha_{p,p'}^a
\widetilde c_{T}^a (p', Z_{T-1}(\omega_i)) \}
$$
and resulting value obtained by each sample path
\begin{equation*}
\widetilde v_{T-1}(p,\omega_i) := r_{T-1}(p,
Z_{T-1}(\omega_i),\pi_{T-1}(p, \omega_i)) + \sum_{p'\in\mathbf{P}}
\alpha_{p,p'}^{\widetilde\pi_{T-1}(p, \omega_i)} \widetilde v_{T}(p',\omega_i)
\end{equation*}
and proceed inductively for $t=T-2,\dots, 1, 0$ until sample
$(\widetilde v_0(p,\omega_i))_{i=1}^n$ is obtained. Many authors
(e.g. \cite{longstaff_schwartz2001}) has shown that the mean of
$(\widetilde v_0(p,\omega_i))_{i=1}^n$ converges in probability to
$v^{\pi^*}_0(p,z_0)$ as the number of sample paths and size of the
regression basis grows to infinity. If the true value function can be
expressed exactly as a linear combination of the selected basis
functions, then the convergence is almost sure when $n\to\infty$
\cite{clement_etal2002}.

\section{Lower and upper bounds}

Now it is clear that a lower bound for $v^{\pi^*}_0(p_0, z_0)$ is
given by
\begin{equation}
  \EE\left[\sum_{t=0}^{T-1} r_t(P_t, Z_t,\widetilde\pi_t(P_t, Z_t)) 
    + \varphi_{t+1}(P_t, Z_t, \widetilde\pi_t(P_t, Z_t)) + r_T(P_T, Z_T)\right]
  \label{lowerBound}
\end{equation}
where $(P_0, Z_0) = (p_0,z_0)$ a.s., $\widetilde\pi$ is some decision
policy, and $(\varphi_t)_{t=1}^T$ are zero mean and independent random
variables. Similarly, an upper bound is given by the expectation of
the following pathwise maximum
\begin{equation}
  \max_{\pi}\sum_{t=0}^{T-1} r_t(P_t, Z_t, \pi_t) + \varphi_{t+1}(P_t, Z_t, \pi_t) + r_T(P_T,Z_T).
  \label{upperBound}
\end{equation}
When $(\varphi_t)_{t=1}^T$ are zero mean and independently
distributed, the upper bound represents the case where the controller
has perfect foresight into the future. It turns out that the careful
choice of $(\varphi_t)_{t=1}^T$ affects the location of these bounds.
It is not hard to see that when $\widetilde \pi_t(x) = \pi^*_t(x)$ and
$\varphi_t(p,z,a)$ is given by
\begin{equation*}
  \sum_{p'\in\PPP} \alpha_{p,p'}^a \left(\EE [v^{{\pi^*}}_{t+1}(p',f_{t+1}(W_{t+1},z))]-v^{{\pi^*}}_{t+1}(p', f_{t+1}(W_{t+1},z)]) \right)
  \label{optMart}
\end{equation*}
for $t=0,\dots,T-1$, both the lower and upper bounds coincide and
gives the value function $v^{\pi^*}_0(p_0,z_0)$.  This can be verified
by substitution into \eqref{lowerBound} and \eqref{upperBound}. Please
see Section 5 in \cite{belomestny_etal2010} for the rigorous details.

In practice, the true value functions $v^{\pi^*}_{t}$ are unknown since its
knowledge vitiates the need to perform numerical work in the first
place. However, the function approximations from the least squares
Monte Carlo can be used in their place instead i.e. 
\begin{equation}
  \sum_{p'\in\PPP} \alpha_{p,p'}^a \left(\frac{1}{I} \sum_{i=1}^I \widetilde v_{t+1}(p',f_{t+1}(W^{(i)}_{t+1},z))- \widetilde v_{t+1}(p',f_{t+1}(W_{t+1},z))\right)
  \label{pracMart}
\end{equation}
for some number $I$ and where $\widetilde v_T(p,z) = r_T(p,z)$ and
$$
\widetilde v_{t+1}(p, z) = \max_{a\in\AAA} r_{t+1}(p,z,a) +
\sum_{p'\in\PPP} \alpha_{p,p'}^a \widetilde c^a_{t+1}(p',z), \quad
t=T-2,\dots,0.
$$
With this substitution, the closer our regression approximations are
to their true counterparts, the tighter the bound estimates and the
smaller their standard errors. In this manner, these bound estimates
allow us to partially gauge the quality of our function approximations
as well as proving bounds for $v_0^{\pi^*}(p_0,z_0)$.

\section{Demonstration: Bermudan put}

The following numerical experiment was run on a Linux Ubuntu 16.04
machine with Intel i5-5300U CPU @2.30GHz and 16GB of RAM using the
author's \emph{R} package \emph{rlsm} which can be found at:
\url{https://github.com/YeeJeremy/rlsm}, and the package manual can be
found at
\url{https://github.com/YeeJeremy/RPackageManuals/blob/master/rlsm-manual.pdf}.
In what follows, a Bermudan put option is considered.  A Bermudan put
option gives the owner the right but not the obligation to sell the
underlying asset for a contracted strike price $K$ at prespecified
time points.  In this setting,
$\mathbf{P}=\{\text{excerised}, \text{unexercised}\}$ and
$\mathbf{A}=\{\text{exercise}, \text{don't exercise}\}$. At $P_t=$
``unexercised'', applying $a=$ ``exercise'' and $a=$ ``don't
exercise'' leads to $P_{t+1}=$ ``exercised'' and $P_{t+1}=$
``unexercised'', respectively with probability one. If $P_t=$
``exercised'', then $P_{t+1}$ = ``exercised'' with probability
one. Now represent the interest rate by $\kappa$ and underlying asset
price by $z$, the reward and scrap functions are given by
\begin{align*}
r_{t}(\text{unexercised}, z, \text{exercise}) &= e^{-\kappa t}(K -  z)^+,\\
r_{T}(\text{unexercised}, z) &= e^{-\kappa T}(K-  z)^+,
\end{align*}
for all $z \in \RR_{+}$ and zero for other $p\in\PPP$ and
$a\in\AAA$. In the above $(K-z)^+ := \max(K-z,0)$. The fair price of
the option is given by
$$ 
v^{\pi^*}_{0}(\text{unexercised}, z_0) = \max\left\{\EE(\max(e^{-\kappa \tau}(K-
  Z_{\tau}), 0) ): \tau = 0,1, \dots, T\right\}.
$$
The option is assumed to reside in the Black-Scholes world
where the asset price process $(Z_{t})_{t=0}^{T}$ follows geometric
Brownian motion i.e.
\begin{equation*}
  Z_{t+1} = e^{(\kappa - \frac{\text{vol}^2}{2})\Delta_{t+1} + \text{vol}\sqrt{\Delta_{t+1}}W_{t+1}} Z_{t}
\end{equation*}
where $(W_{t})_{t=1}^{T}$ are independent standard normal random
variables, $\Delta_{t+1}$ is the time step and $\text{vol}$ is the
volatility of stock returns. 

Let us set up our model in the below code listing. In this example,
the package \emph{StochasticProcess}\cite{StochasticProcess} was used
to generate our paths. However, the user is free to do so in which
ever manner they wish to.  The simulated paths are represented by
object \texttt{path} which gives a 3 dimensional array where entry
$[i,j,k]$ gives the $j$-th component of $Z_{k-1}(\omega_i)$. For the
case that $P_t$ is governed deterministically by the actions, users
can specify a control matrix (Line 10) instead of the more tedious
transition probabilities (Lines 11-15).

\begin{lstlisting}[caption = {Set up}]
library(StochasticProcess)
## Parameters
set.seed(123)
step <- 0.02  # Step size
kappa <- 0.06 * step  ## Adjust interest according to step size
vol <- 0.2 * sqrt(step) ## Adjust vol according to step size
n_dec <- 51  # Number of decision times T + 1
strike <- 40  # Strike price
## The transition for P_t. See manual for more information.
control <- matrix(c(c(1, 1), c(2, 1)), nrow = 2, byrow = TRUE)
## control <- array(data = 0, dim = c(2,2,2))
## control[2,1,2] <- 1
## control[2,2,1] <- 1
## control[1,1,1] <- 1
## control[1,2,1] <- 1
## Reward and scrap functions
Reward <- function(state, time) {
    output <- array(data = 0, dim = c(nrow(state), 2, 2))
    output[, 2, 2] <- exp(-kappa * (time - 1)) * pmax(strike - state, 0)
    return(output)
}
Scrap <- function(state) {
    output <- matrix(data = 0, nrow = nrow(state), ncol = 2)
    output[, 2] <- exp(-kappa * (n_dec - 1)) * pmax(strike - state, 0)
    return(output)
}
## Simulate paths to do regression on
n_path <- 10000  # Number of paths
start <- 36  # Starting state Z_0
path <- GBM(start, kappa, vol, n_dec, n_path, TRUE)  # Generated paths
\end{lstlisting}

\subsection{Choice of basis functions}

It is well known that the quality of the LSM results depend on an
appropriate choice of the regression basis. With this in mind, the
\emph{rlsm} package aims to allow users to specify any possible set of
basis functions using a combination of the following six objects. Not
all objects are required but atleast one from Lines 31, 33, 34 or 35
must be supplied. Please keep in mind that some \emph{R} functions
from the package will have default values for these parameters.

\begin{lstlisting}[caption = {Regression basis}]
basis <- matrix(c(1, 1), nrow = 1)
btype <- "power"  # currently either "power" or "laguerre"
intercept <- TRUE
knots <- matrix(c(30, 40, 50), nrow = 1)
BasisFunc <- function(state) { 1 / state }
n_rbasis <- 1
\end{lstlisting}

Suppose we are performing the regression at time $t$.  
\begin{itemize}
\item The first object \texttt{basis} describes some transformation of
  the components of $Z_t = [Z_t^{(1)},\dots,Z_t^{(d)}]^T$. If
  \texttt{btype=''power''} and if entry $[i,j]$ is non-zero, then
  $(Z_t^{(i)})^j$ is included in the regression basis. If
  \texttt{btype=''laguerre''} and if entry $[i,j]$ is non-zero, then
  the j-th Laguerre polynomial of $Z_t^{(i)}$ is included in the
  regression basis. The object \texttt{basis} is processed row-wise.
\item The object \texttt{intercept} decides whether a constant ($1$)
  is added to the regression basis.
\item The object \texttt{knots} gives the location of the knots
  used for linear splines. If entry $[i,j]$ is given by $B$, then
  $(Z^{(i)}_t-B)^+$ is added to the basis.  The object \texttt{knots}
  is processed row-wise.
\item The object \texttt{BasisFunc} is a user defined function which
  which acts on an $n\times d$ matrix representing the
  $(Z_t(\omega_i))_{i=1}^d$ where entry $[i,j]$ gives
  $Z^{(j)}_t(\omega_i)$. This function should output a matrix to
  append to the design matrix horizontally on the right. The object
  \texttt{n\_rbasis} gives the number of basis functions added by the
  \texttt{BasisFunc} function and must be supplied if
  \texttt{BasisFunc} is used.
\end{itemize}

The order in which the objects are processed is \texttt{basis},
\texttt{intercept}, \texttt{knots}, and \texttt{BasisFunc}.  So in
Listing 2, the regression basis is set to be
$\{Z_t, Z_t^2, 1, (Z_t-30)^+, (Z_t-40)^+, (Z_t-50)^+, 1/Z_t \}$ and in
that order.

\subsection{Choice of regression}

Recall that the default method used in the $lm$ function is QR
factorization with pivoting. However, the default regression method
used in the \texttt{LSM()} function is linear least squares using
singular value decomposition (SVD) taking into account any rank
deficiency in the design matrix. We do this for the following
reasons. First, the least squares Monte Carlo method typically
generate sample paths from a single point at the start. Therefore, as
we perform regressions closer to $t=0$, the design matrix is more
likely to be rank deficient and so the SVD method is more stable than
the QR approach. Secondly, when the number of rows in the design
matrix is drastically larger than the number of columns, there is very
little difference in the computational effort between SVD and QR. This
is often the case for least squares Monte Carlo where the number of
sample paths are subtantially larger than the size of the regression
basis. We point the reader to Section 3 in \cite{demmelBook} for a
detailed discussion.

\begin{lstlisting}[caption = {Regression}]
## SVD
lsm1 <- LSM(path, Reward, Scrap, control, basis, intercept, btype, TRUE, knots, BasisFunc, n_rbasis)
## QR factorization
RegFunc <- function(x, y, tt) {
    out <- array(lm(y~ 0 + x)$coefficients)
    out[is.na(out)] <- 0
    return(out)
}
lsm2 <- LSM(path, Reward, Scrap, control, basis, intercept, btype, TRUE, knots, BasisFunc, n_rbasis, Reg = RegFunc)
\end{lstlisting}

In the above we test the default SVD regression approach with the QR
approach from the base \texttt{lm()} function in \emph{R} and we get
the same results as shown below. However, in our experiments, adding
$\{Z_t^3,Z_t^4\}$ to the regression basis causes SVD and QR to give
different resuts. This is due to how they differ in the way they
handle rank deficiency in the design matrix.

\begin{lstlisting}[caption = {Value estimates}]
> print(mean(lsm2$value[,2,1]))
[1] 4.468097
> print(mean(lsm1$value[,2,1]))
[1] 4.468097
\end{lstlisting}

In Listing 4, the \texttt{RegFunc} allows the user to specify any
function from \emph{R} to use in least squares Monte Carlo. The only
condition is that it should return real valued coefficients for each
of the feature functions in the regression basis. This is why we
convert any \texttt{NA} values to $0$ on Line 42. This is useful
considering the large amount of statistical tools coded in \emph{R}.

\subsection{Lower and upper bounds}

Let us finally demonstrate the construction of the upper and lower
bounds. Line 54 extracts the prescribed policy using
\eqref{bellman}. Line 55 computes \eqref{pracMart}. Line 56 computes
the lower and upper bound estimates in Section 4. Note that we use the
\texttt{NestedGBM()} function from the \emph{StochasticProcess}
package to generate the nested simulation in \eqref{pracMart} but the
user is free to generate it in anyway they want to. The object
\texttt{subsim} should be a 4 dimensional array where entry
$[i,j,k,l]$ represents the $j$-th component of
$f_{l}(W^{(i)}_l,Z_{l-1}(\omega_i))$. For the case that $j=1$, we can
represent it as a 3 dimensional array $[i,k,l]$ instead as done below.

\begin{lstlisting}[caption = {Lower and upper bounds}]
n_path2 <- 100
path2 <- GBM(start, kappa, vol, n_dec, n_path2, TRUE)
n_subsim <- 100  ## Number of nested simulations I
subsim <- NestedGBM(path2, kappa, vol, n_subsim, TRUE)  # nested simulations
policy <- PathPolicy(path2, lsm1$expected, Reward, control, basis, btype, TRUE, knots, BasisFunc, n_rbasis) # Prescribed policy
mart <- AddDual(path2, subsim, lsm1$expected, Reward, Scrap, control, basis, btype, TRUE, knots, BasisFunc, n_rbasis)  # varphi
bounds <- Bounds(path2, Reward, Scrap, control, mart, policy)
\end{lstlisting}

The below then generates the $99\%$ confidence interval for the fair
price of the option using the function approximations from LSM.

\begin{lstlisting}[caption = {99\% confidence intervals}]
> print(GetBounds(bounds, 0.01, 2))
[1] 4.361008 4.567014
\end{lstlisting}

\section{Final thoughts}

Let us finally discuss the computational times. It takes around 0.35
cpu seconds for the \texttt{LSM()} on Line 38 to run and the same
amount of time to compute both the lower and upper bounds on Lines 55
and 56. This package provides an easy way for users of least squares
Monte Carlo to experiment with the large amount of statistical tools
developed by the \emph{R} community.

\bibliography{main}
\bibliographystyle{amsplain}

\end{document}